\pdfoutput=1

\documentclass[aps,prl,twocolumn,showpacs,amsmath,amssymb,amsfonts,long]{revtex4}
\usepackage{graphicx}

\newcommand{\mysection}[2]{\textbf{#1} \textsl{#2}. --}

\begin{document}

\title{Self-refraction of supernova neutrinos: mixed spectra and three-flavor instabilities}

\author{Alexander Friedland} \affiliation{Theoretical Division, MS B285, Los
Alamos
  National Laboratory, Los Alamos, NM 87545, USA}

\date{January 6, 2010; updated to match the PRL version on May 10, 2010}

\begin{abstract}
Neutrinos in a core-collapse supernova undergo coherent flavor transformations in their own background.
We explore this phenomenon during the cooling stage of the explosion. Our three-flavor calculations reveal qualitatively new effects compared to a two-flavor analysis. These effects are especially clearly seen for the inverted mass hierarchy: we find a different pattern of spectral ``swaps'' in the neutrino spectrum and a novel ``mixed'' spectrum for the antineutrinos. A brief discussion of the relevant physics is presented, including the instability of the two-flavor evolution trajectory, the 3-flavor pattern of spectral ``swaps,'' and partial nonadiabaticity of the evolution.

\end{abstract}


\maketitle

\mysection{1.}{Introduction and motivation}  
Core-collapse supernovae play an essential role in the evolution of the Universe, from controlling the temperature of the gas and the rate of star formation in the galactic disk ({\it e.g.}, \cite{Scannapieco:2008vf}), to synthesizing and dispersing heavy elements ({\it e.g.}, \cite{Qian:2007vq}). 
The ashes of ancient explosions have literally shaped the world around us. 
The struggle to understand and model the explosion mechanism has occupied researchers for the better part of the 20th century \cite{Baade:wo,Burbidge:1957vc,Colgate:1960zz,Colgate:1966ax,Freedman:1973yd,Arnett:1977xj,Bethe:1979zd,Bowers:ea,Bethe:1984ux,Wilson:1986ha} and remains a very active topic (for review, see, {\it e.g.}, \cite{Bethe:1990mw,Woosley:ao}).

Neutrinos, which are emitted during the first $\sim10$ sec after the onset of the collapse and arrive to us directly from the core of the star, could serve as a direct probe of the explosion mechanism. It is expected that, unlike supernova 1987a \cite{Bionta:1987qt,Hirata:1987hu},
the next \emph{galactic} supernova may yield $O(10^{4})$ antineutrino ($\bar\nu_{e}$) events and, furthermore, a similar number of  neutrino ($\nu_{e}$) events, if a large liquid argon detector is constructed at DUSEL. With such high statistics, it should be possible to reconstruct the second-by-second evolution of the $\nu_{e}$ and $\bar\nu_{e}$ spectra.  
The task is to understand how to ``read'' this signal, {\it i.e.}, how to extract signatures of various physical processes from it, and to optimize the detector design.

One important process that needs to be thoroughly understood is the coherent flavor transformations of neutrinos outside of the neutrinosphere. Compared to the well-studied case of solar neutrinos, in a supernova these transformations are much more involved. Not only neutrinos and antineutrinos of all flavors are emitted, not only are there two mass splittings -- ``solar'' and ``atmospheric'' -- to worry about \cite{Dighe:1999bi}, but the physics of the transformations is significantly richer. For example, several seconds after the onset of the explosion, the flavor conversion probability is affected by the expanding shock front \cite{Schirato:2002tg} and the turbulent region behind it \cite{Friedland:2006ta}. The conversion process in such a ``bumpy,'' stochastic profile is qualitatively different from the adiabatic MSW effect in the smooth, fixed  density profile of the Sun. 

Even more complexity is brought about by the coherent scattering of neutrinos off each other \cite{Fuller:wm,Notzold:1987ik,Pantaleone:1992xh,Pantaleone:1992eq,Sigl:1992fn,McKellar:1992ja,Samuel:1993uw,Kostelecky:1993dm,Pantaleone:1994ns,Kostelecky:1994dt,Samuel:1995ri,Pastor:2001iu,Pastor:2002we,Friedland:2003dv,Bell:2003mg,Friedland:2003eh,Friedland:2006ke}. This neutrino ``self-refraction'' 
 results in highly nontrivial flavor transformations \cite{Duan:2005cp,Hannestad:2006nj,Duan:2006an,Duan:2006jv,Raffelt:2007cb,Raffelt:2007xt,Fogli:2007bk,Duan:2007sh,EstebanPretel:2007yq,Dasgupta:2008cd,EstebanPretel:2008ni,Dasgupta:2009mg} close to the neutrinosphere, typically within a few hundred kilometers from the center, where the density of streaming neutrinos is very high. Since the evolving flavor composition of the neutrino flux feeds back into the oscillation Hamiltonian, the problem is \emph{nonlinear}. Furthermore, as the interactions couple neutrinos and antineutrinos of different flavors and energies, the oscillations are characterized by \emph{collective} modes. This leads to very rich physics that has been the subject of intense interest over the last several years. 
 
One may wonder whether all this complexity will impede the extraction of useful information from the future signal. In fact, the opposite is true: the new effects can \emph{imprint} information about the inner workings of the explosion on the signal. For example, by observing the signatures of the expanding shock and the post-shock region in the neutrino signal, we will learn about the development of the explosion during the crucial first 10 seconds. This information could be inaccessible in other ways. 

On the other hand, it is fair to say that the subject is still far from being exhausted and qualitatively new effects continue to be uncovered.  This letter is another contribution to this effort. We explore the self-refraction phenomenon for the conditions typically present during the cooling stage of the protoneutron star (several seconds into the explosion) and show that a three-flavor analysis of this process reveals several new effects.
Below we report the main results; a detailed discussion will be given elsewhere \cite{inprep}.


\begin{figure*}[t]
  \centering
  \includegraphics[angle=0,width=0.97\textwidth]{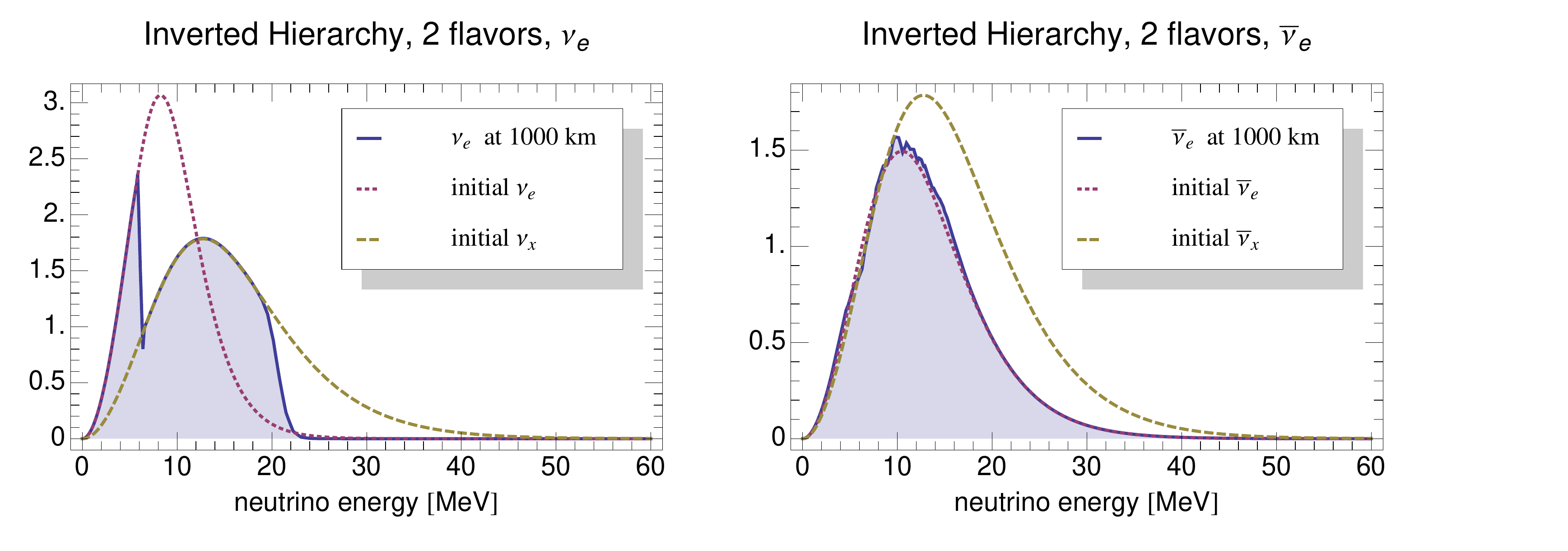}
    \includegraphics[angle=0,width=0.97\textwidth]{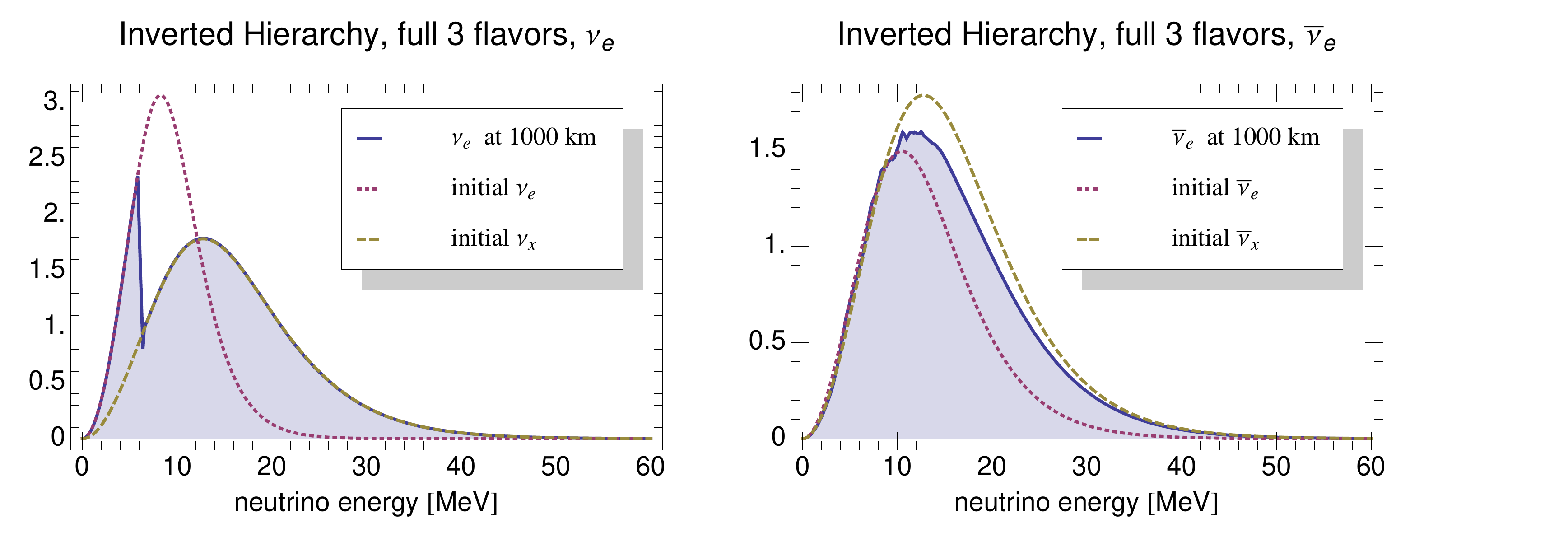}
   \caption{Spectra of $\nu_{e}$ \emph{(left)} and $\bar\nu_{e}$ \emph{(right)} at 1000 km, shown by the filled regions. The top plots are computed within the two-flavor framework, the bottom ones with a full three-flavor calculation. Inverted mass hierarchy and $\theta_{13}=0.01$ are assumed. Also shown, by dashed/dotted curves, are the original spectra at the neutrinosphere, as labeled.}
  \label{fig:spectra2flavors}
  \end{figure*}

\mysection{2.}{Formulation of the problem} The energy spectra of emitted neutrinos evolve as the explosion develops. 
Since the self-refraction effect is nonlinear, different initial spectra may result in qualitatively different outcomes. Therefore, it is important to choose spectra that are (i) specific to the late phase and (ii) simulated with sufficient accuracy. It has been shown \cite{Keil:2002in} that the emerging spectra of the nonelectron neutrinos ($\nu_{x}$) are sensitive to several seemingly subdominant processes, such as $\nu_{x}e^{\pm}$ scattering, neutrino coannihilation, bremsstrahlung, and nucleon recoil. The addition of nucleon recoil, in particular, makes the $\nu_{x}$ spectra softer and brighter, changing the average energy from $\sim26$ MeV to $\sim19$ MeV \cite{Keil:2002in} \footnote{We have also verified this with our simplified model of neutrino decoupling.}. This significant shift may alter the collective oscillations, and needs to be included. 

Indeed, collective oscillations with such spectra were recently studied in \cite{Dasgupta:2009mg} and several interesting features were observed. In particular, for the inverted hierarchy (IH), the neutrino spectra were found to be exchanged between $\nu_{e}$ and $\nu_{x}$, but only in an interval of energies, from about $6$ to $22$ MeV. The neutrinos outside of this interval were unchanged, resulting in the presence of \emph{two spectral splits}. While spectral splits had been observed with other fluxes, starting with the seminal work \cite{Duan:2006an}, and in fact seem quite ubiquitous,  \emph{multiple} splits is a new and interesting phenomenon, with potentially important implications for signal detection (high-energy splits would be easy to detect).

While the analysis in \cite{Dasgupta:2009mg} is an important step toward understanding the collective oscillations of late-time neutrinos, it was limited to two flavors. What happens with all three flavors included in the calculation? {\it A priori}, several possibilities come to mind. The effects of the third state could be a small correction, as in the cases of solar, KamLAND, and atmospheric neutrino oscillations. Alternatively, the effects of the two splittings could be large, but factorizable, just like the conventional MSW effects in a supernova, which can be treated ``pairwise'' between the states and combined at the last step \cite{Dighe:1999bi}. Instead, something even more interesting happens: the three-flavor evolution with a nonzero solar mass splitting gives an entirely new result, as we display next.

\mysection{3.}{Setup of the calculation} 
For the late-time spectra, we use the results of the Monte-Carlo simulations from \cite{Keil:2002in}. For the calculations shown in this letter, we select the point $p=10$, $q=3.5$ from Table 6 in that paper, corresponding to the spectrum emitted from a steep power-law profile near the neutrinosphere ($\rho\propto r^{-p}$, $T\propto r^{-q}$). For the matter profile at $r\sim100-1000$ km we assume a neutrino driven wind with $\rho=\rho_{0}(10\mbox{ km}/r)^{3}$. We take $\rho_{0}=2\times10^{9}$ g/cm$^{-3}$, and $Y_{e}= 0.5$. 

Our three-flavor calculation is carried out with the following parameters: $|\Delta m_{\rm atm}^{2}|=2.7\times10^{-3}$ eV$^{2}$, $\Delta m_\odot^{2}=7.7\times 10^{-5}$ eV$^{2}$, $\theta_{13}=0.01$, and $\sin^{2}\theta_{12}=0.31$. 
In the two-flavor calculation, we set the solar mixing angle $\theta_{12}$ to zero and drop the state that in vacuum is separated from the predominately $\nu_{e}$ ($\bar\nu_{e}$) state by $\Delta m_\odot^{2}$.

We perform multienergy, single-angle calculations of the evolution, starting at 40 km and ending at 1000 km.

\mysection{4.}{Results: comparison of two- and three-flavor runs}
The resulting spectra at 1000 km for the IH case ($\Delta m_{\rm atm}^{2}<0$) are presented in Fig.~\ref{fig:spectra2flavors}. We concentrate here on this case, because it most clearly illustrates the effect; both hierarchies are treated and contrasted in \cite{inprep}. The top panels show the two-flavor calculations, the bottom ones, the corresponding three-flavor runs. The $\nu_{e}$ spectra are on the left, those for $\bar\nu_{e}$ are on the right. The dashed curves and the dotted curves show the corresponding initial spectra (see legend). Animations showing the evolution of the spectra as a function of the radius for both IH and NH are available as Supplemental Material online \cite{website,PRL_website}.

The results of the two-flavor calculations appear to be in very good agreement with the inverted hierarchy calculations of \cite{Dasgupta:2009mg}. 
The three-flavor calculation results are visibly different: (i) the high-energy split in the $\nu_{e}$ channel is gone; (ii) in the $\bar\nu_{e}$ channel, the flavor swap probability is neither zero, nor one, but increases \emph{gradually} with neutrino energy. The swap is \emph{incomplete} in this case.

\begin{figure}[t]
  \includegraphics[angle=0,width=0.5\textwidth]{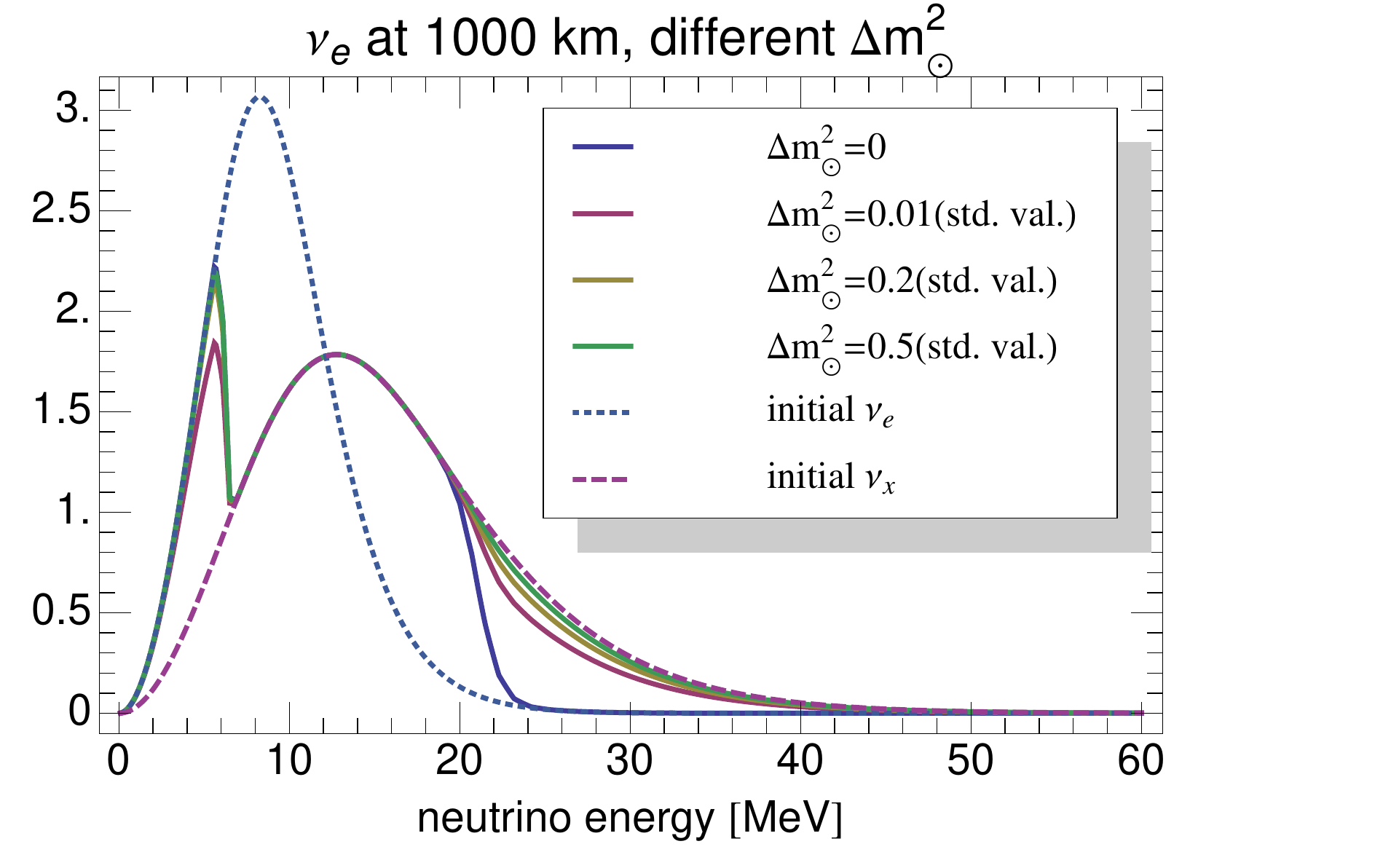}
  \caption{Impact of decreasing the solar mass splitting, $\Delta m_{\odot}^{2}$, on the neutrino spectra at 1000 km. Notice that, while a strictly vanishing $\Delta m_{\odot}^{2}$ gives the two-flavor result, even a tiny nonzero value of $\Delta m_{\odot}^{2}$ qualitatively changes the answer.}
  \label{fig:nuedecreasesolar}
\end{figure}

\begin{figure}[t]
  \includegraphics[angle=0,width=0.5\textwidth]{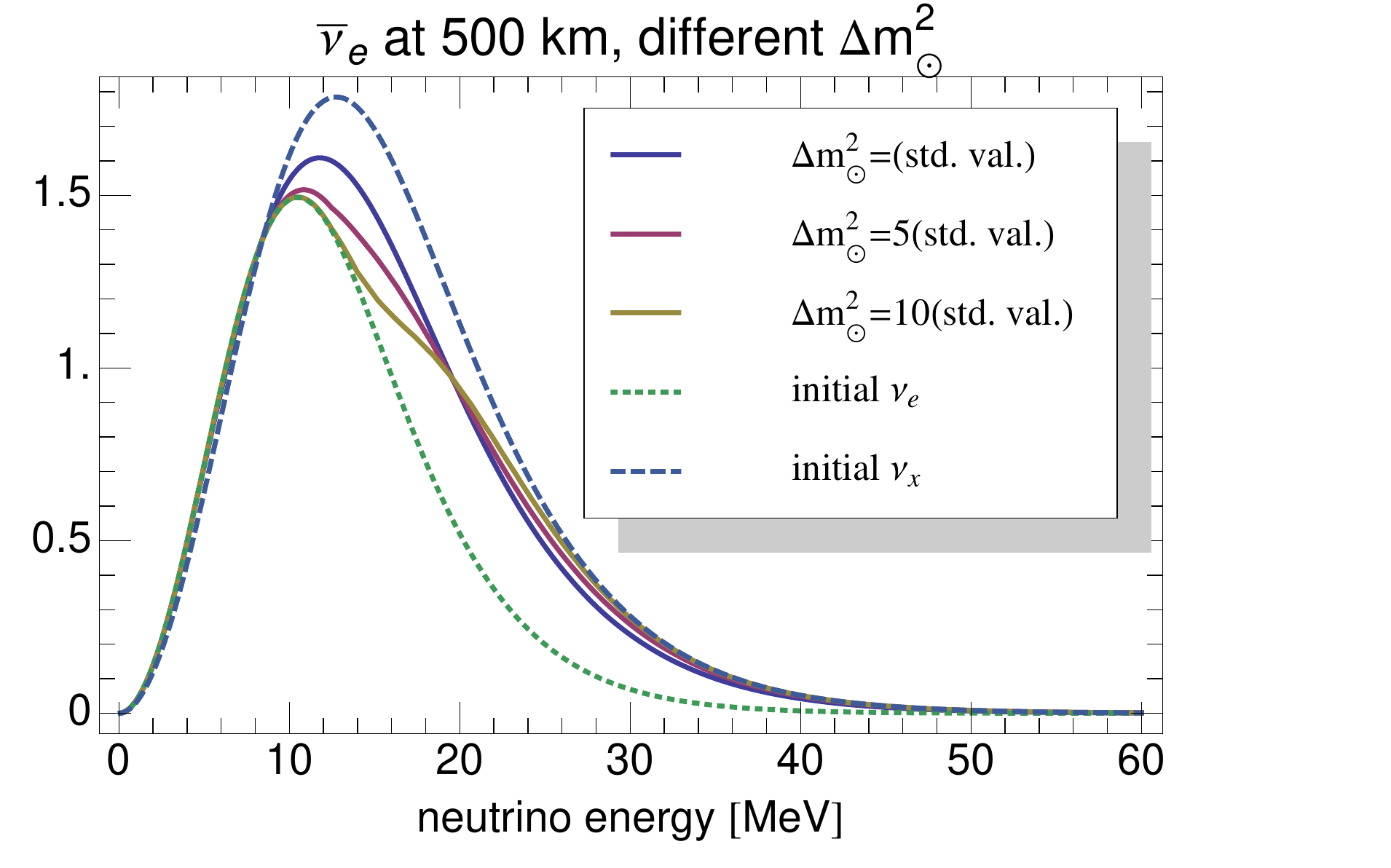}
  \caption{Impact of increasing the solar mass splitting $\Delta m_{\odot}^{2}$ on the neutrino spectra at 500 km. Large $\Delta m_{\odot}^{2}$ makes the evolution more adiabatic, leading to the formation of a (broad) spectral split. }
  \label{fig:nuebarincreasesolar}
\end{figure}

\mysection{5.}{Discussion}  
Both of these results deserve further investigation. 
First we can establish that the physics driving flavor conversion before 1000 km are the neutrino-neutrino interactions, with the conventional MSW effect being unimportant. The atmospheric level crossing does occur here, but for the chosen parameters it is strongly nonadiabatic (flavor preserving). Moreover, it occurs when $r\gtrsim600$ km, by which point the neutrino self-refraction effects have ceased. The small MSW effects are seen in the $\bar\nu_{e}$ channel as small wiggles.

It looks paradoxical at first that the solar splitting completely changes the evolution of neutrinos at high energies, even though it is only $\sim3$\% of the atmospheric splitting. Clearly, at this level $\Delta m_{\odot}^{2}$ cannot ``overpower'' $\Delta m_{\rm atm}^{2}$; what it can do, however, is change the nature of the collective motion, allowing it to extend from the 2-flavor subspace into the full 3-flavor space. Evidently, this results in a radical change, not a small correction.

Let us investigate how this change happens as the value of the solar splitting is turned on. The results are shown in Fig.~\ref{fig:nuedecreasesolar}. These at first may be even more surprising: the two-flavor spectrum is reproduced only  when $\Delta m_{\odot}^{2}=0$; as soon as it is nonzero, even very small, the high-energy split disappears. Since for $\Delta m_{\odot}^{2}=7.7\times10^{-7}$ eV$^{2}$ (1\% of its true value) the corresponding oscillation length is $10^{4}$ km -- much longer than the scales in the problem -- one might think the two-flavor limit should be reached. Instead, the spectrum in this case is closer to the realistic three-flavor one than to the two-flavor one.

To understand this result, it proves useful to examine the evolution as a function of radius in the matter mass basis \cite{website,PRL_website,inprep}. Neutrinos, initially in the mass eigenstates, develop an instability which leads to large collective oscillations\cite{Kostelecky:1993dm,Duan:2005cp,Hannestad:2006nj}. What is interesting in our case is that, shortly after the collective oscillations develop between the ``atmospheric'' eigenstates ($\nu_{2,3}$), the third state ($\nu_{1}$) joins in. Therefore, not only is the initial configuration unstable, but \emph{the two-flavor trajectory is also unstable} in the three-flavor space. A small nonzero $\Delta m_{\odot}^{2}$ is enough to displace the system from the two-flavor subspace; the instability makes it then run away from it (driven primarily by $\Delta m_{\rm atm}^{2}$). 
The role of small $\Delta m_{\odot}^{2}$ is thus similar to that of small $\theta_{13}$ in the development of the Kostelecky-Samuel instability \cite{Kostelecky:1993dm,Duan:2005cp,Hannestad:2006nj}.

\begin{figure*}[t]
  \includegraphics[angle=0,width=0.45\textwidth]{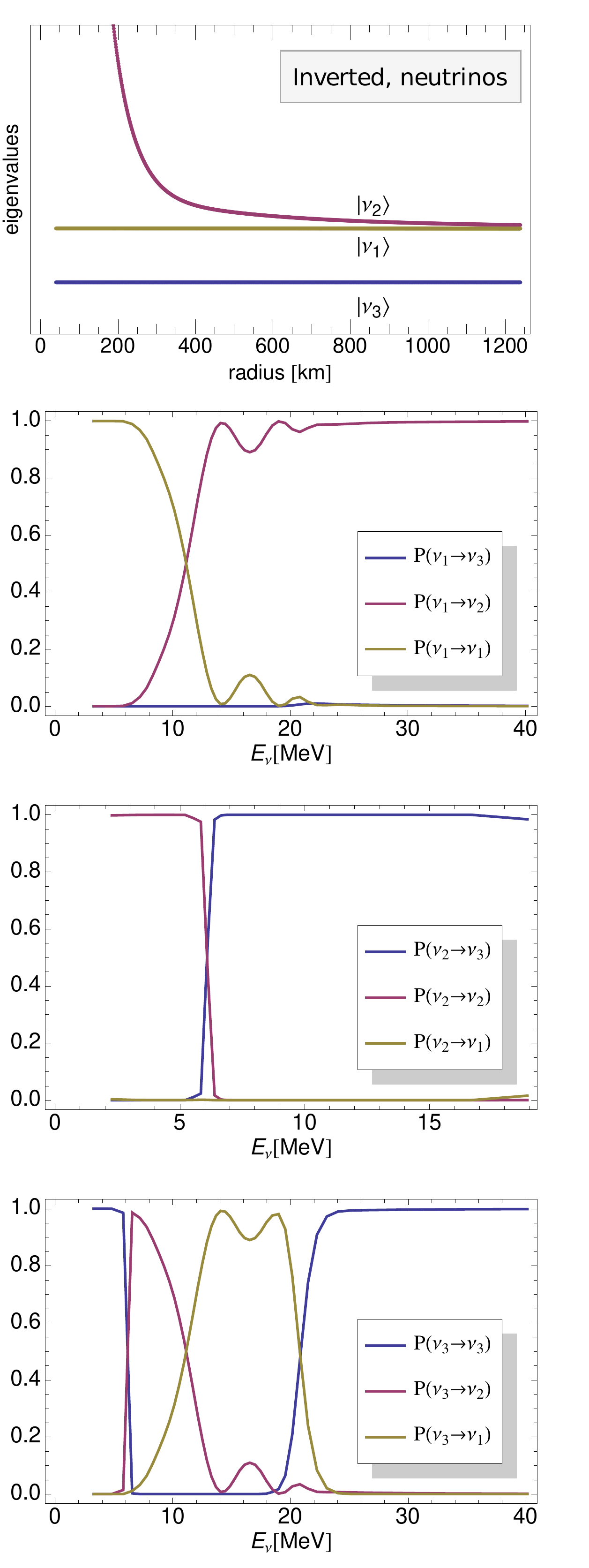}
  \caption{Final three-flavor conversion probabilities for neutrinos, shown in the mass basis. For example, $P(\nu_{1}\rightarrow\nu_{2})$ indicates the probability, as a function of energy, that the neutrino originally in eigenstate $|\nu_{1}\rangle$ transitions into eigenstate $|\nu_{2}\rangle$ by the end of the collective transformations (at 1000 km). The inverted mass hierarchy is assumed. The top panel specifies the labeling convention for the states. (This plot appears in Supplemental Online Material in the Phys. Rev. Lett. version, \cite{PRL_website}.)}
  \label{fig:splits_mattermassbasis_nu}
\end{figure*}

\begin{figure*}[t]
  \includegraphics[angle=0,width=0.47\textwidth]{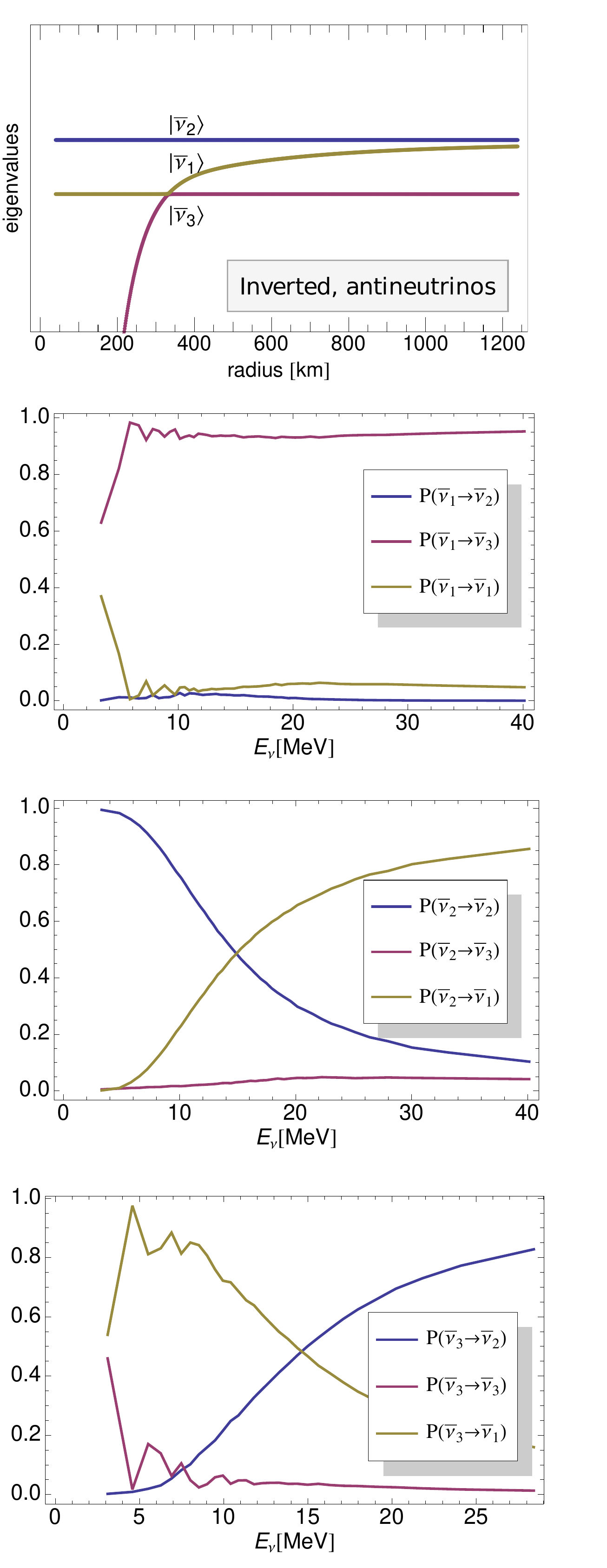}
  \caption{Same as Fig.~{\protect\ref{fig:splits_mattermassbasis_nu}}, but for the antineutrinos. (This plot appears in Supplemental Online Material in the Phys. Rev. Lett. version, \cite{PRL_website}.)}
  \label{fig:splits_mattermassbasis_nubar}
\end{figure*}

Since the evolution is no longer constrained to the two-flavor subspace, its final state may be expected to be different. 
Indeed, it is.
In the two-flavor case, the final state has a swap $\nu_{2}\leftrightarrow\nu_{3}$ on the interval $6 \lesssim E_{\nu}\lesssim22$ MeV\footnote{Here $\nu_{3}$ is the lower-energy state, following the standard notation for the IH.}. In contrast, the final state of the three-flavor calculation is significantly more complicated, as illustrated in Figs.~\ref{fig:splits_mattermassbasis_nu} and \ref{fig:splits_mattermassbasis_nubar} \footnote{These two figures appear in Phys. Rev. Lett. as supplemental online material, \cite{PRL_website}, because of space constraints.}. The pattern is the same only up to  $E_{\nu}\sim10$ MeV, with the third state $\nu_{1}$ being a spectator, $\nu_{1}\rightarrow\nu_{1}$. For  $10 \lesssim E_{\nu}\lesssim20$ MeV we instead observe \emph{a cyclical permutation}: $\nu_{2}\rightarrow\nu_{3}$, $\nu_{3}\rightarrow\nu_{1}$, $\nu_{1}\rightarrow\nu_{2}$; while for  $E_{\nu}\gtrsim20$ MeV it is the $\nu_{3}$ state that is a spectator and the other two swap spectra:  $\nu_{1}\leftrightarrow\nu_{2}$, $\nu_{3}\rightarrow\nu_{3}$. Since the initial spectra of $\nu_{1}$ and $\nu_{3}$ are identical, no second split is seen at 1000 km. See \cite{inprep} for more details.

Figure~\ref{fig:nuedecreasesolar} also shows another important role of $\Delta m_{\odot}^{2}$: as it nears its physical value, the spectrum at high energy becomes closer to the emitted $\nu_{x}$ spectrum. This means the neutrinos at the end of the collective oscillations are ``put'' into the Hamiltonian eigenstates. This happens when the evolution is adiabatic \cite{Duan:2006an,Raffelt:2007cb,Raffelt:2007xt}. Observe that adiabaticity is broken even when $\Delta m_{\odot}^{2}$ is 20-50\% of its physical values. This means that adiabaticity for physical $\Delta m_{\odot}^{2}$ is only marginal in this channel.

Indeed, let us compare the neutrino vacuum oscillation length to the scale height of the neutrino-neutrino potential. The latter is a power law $\propto r^{-4}$, so the scale height is $|d\ln H_{\nu\nu}/dr|^{-1}\sim r/4 \sim 75-100$ km for $r\sim300-400$ km. The atmospheric splitting for $E\sim 15$ MeV gives a characteristic scale $2E/\Delta m_{\rm atm}^{2}\sim 2$ km, so a high degree of adiabaticity is expected (and seen for the low-energy $\nu_{e}$ split). In contrast, for the solar splitting, $2E/\Delta m_{\odot}^{2}\sim 77$ km, so the evolution is only borderline adiabatic \footnote{This order-of-magnitude estimate omits $O(1)$ factors. A more detailed analysis will be presented elsewhere.}.

The weakness of adiabaticity also helps to understand the $\bar\nu_{e}$ spectra: the antineutrinos are not placed into mass eigenstates. 
To increase adiabaticity, one can artificially increase the value of $\Delta m_{\odot}^{2}$. We reran the calculation with $\Delta m_{\odot}^{2}$ 5 and 10 times larger than its actual value. The resulting spectra, shown in Fig.~\ref{fig:nuebarincreasesolar}, indeed exhibited a more ``conventional'' split, centered around $E_{\bar\nu}\sim19$ MeV, although still fairly broad ($\sim5$ MeV and $\sim3$ MeV half-widths correspondingly). The mixed $\bar\nu_{e}$ spectrum found for physical $\Delta m_{\odot}^{2}$ can be thought of as an extremely broad split, with a width comparable to the entire range of the antineutrino energies. (The width of the split is related to the degree of adiabaticity, \cite{Raffelt:2007xt}.)


\mysection{6.}{Conclusions}
The late-time spectra provide an interesting physical system for studying collective transformations. Two- and three-flavor calculations of the phenomenon yield qualitatively different results. The two-flavor trajectory is unstable to small displacements in the three-flavor space. As we saw, even a tiny nonzero value of  $\Delta m_{\odot}^{2}$ is enough to give a different $\nu_{e}$ spectrum. Moreover, the adiabaticity of the $\Delta m_{\odot}^{2}$-driven evolution is marginal, leading to a mixed spectrum in the $\bar\nu_{e}$ channel, rather than the usual complete swap. Clearly, a 3-flavor analysis is a must at this stage of the explosion.


This letter is meant to open, rather than close, the issue. It is also important to consider what happens in the NH case. Briefly, the high-energy splits at $\approx 20$ 
MeV remain, but mixed spectra appear below \cite{inprep}. One should also investigate the sensitivity of the answer to the details of the spectra and what happens in full multiangle calculations. Finally, numerous applications await exploration: how the collective oscillations affect the signatures of the shock and turbulence, whether they change the $r$-process and the diffuse supernova neutrino background, {\it etc}.

We concur with the conclusions of \cite{Dasgupta:2009mg}: the physics of supernova neutrino conversion continues to surprise us with its richness.

\begin{acknowledgments}

I thank J.~Carlson, V.~Cirigliano, and S.~Reddy for feedback. I am especially grateful to H. Duan for numerous helpful discussions and for bringing to my attention Refs.~\cite{Kostelecky:1993dm} and \cite{Dasgupta:2009mg}. This work was supported by the LDRD program of the Los Alamos National Laboratory.

\end{acknowledgments}

\bibliography{latesplits}

\end{document}